# VARIATIONAL STUDY OF LIGHT HYPERNUCLEI

*Bhupali Sharma[1]*

[1] *Department of Physics, Arya Vidyapeeth College, Guwahati, India*
*Email : bhupalisharma@gmail.com*

The real nature of nucleon-hyperon and hyperon-hyperon interaction is best understood by theoretical study of single lambda hypernuclei. Variational Monte-Carlo calculations have been done on some light lambda hypernuclei using realistic NN, NNN ; phenomenological ΛN and ΛNN interactions. The results obtained for different interactions show the role of ΛN and ΛNN interactions. These interactions have been found to be important for the stability of the hypernuclear systems.

PACS numbers : 21.80.+a;21.60.GX;21.45.-v;21.30.-x

## I.  INTRODUCTION :

Since the first report of hypernuclear event by M. Danysz and J. Pniewski[1] in 1953, Hypernuclear Physics has made progress in both experimental as well as theoretical fronts. Since then various confirmed hypernuclear events have been reported experimentally and lots of theoretical studies have been done on hypernuclear field. Different potential models have been proposed and established through theoretical studies over the years and it is still progressing. Theoretical study of lambda hypernuclei is important to understand the real nature of nucleon-hyperon and hyperon-hyperon interaction.

In our earlier studies[2,3] we have done variational Monte-carlo studies on different hypernuclear systems with various potential models in order to predict the possibility of existence of different hypernuclear events theoretically and to study the effect of different parameters of the potential models on the binding energy of the hypernuclear systems.

In the present study we study the role of ΛN and ΛNN interaction parameter and the effect of exchange part of ΛN interaction on the stability of few light hypernuclear systems. From the study we find that that stability of a hypernuclear system depends on two-body ΛN and three-body ΛNN potential. The role of exchange part of ΛNN potential is found to be crucial.

## II.  HAMILTONIAN AND WAVEFUNCTION :

For the nuclear part of the Hamiltonian in the present study, *Argonne $V_{18}$* NN and *Urbana IX* NNN potential same as ref.[2] have been employed .

The phenomenological ΛN potential consists of central, Majorana space-exchange and spin-spin ΛN components and is given by,

$$V_{\Lambda N} = (V_c(r) - \bar{V} T_\pi^2(r))(1 - \varepsilon + \varepsilon P_x) + \frac{1}{4} V_\sigma T_\pi^2(r) \boldsymbol{\sigma}_\Lambda \cdot \boldsymbol{\sigma}_N \qquad (1)$$

where $P_x$ is the majorana space-exchange operator and ε is the space exchange parameter which is taken as 0.2[4]. $V_c(r)$, $\bar{V}$ and $V_\sigma$ are respectively Wood-saxon core, spin-average and spin-dependent strength and $T_\pi^2(r)$ is one-pion tensor shape factor.

The ΛNN potential consists of two terms. Firstly, a two-pion exchange and a dispersive part[5].

The two-pion exchange part of the interaction is given by

$$W_p = -\tfrac{1}{6} C_p (\tau_i \cdot \tau_j)\{X_{i\Lambda} \cdot X_{j\Lambda}\} Y_\pi(r_{i\Lambda}) Y_\pi(r_{j\Lambda}) \quad (2)$$

Where $X_{k\Lambda}$ is the one-pion exchange operator given by,

$$X_{k\Lambda} = (\sigma_k \cdot \sigma_\Lambda) + S_{k\Lambda}(r_{k\Lambda}) T_\pi(r_{k\Lambda}) \text{ with}$$

$$S_{k\Lambda}(r_{k\Lambda}) = 3(\sigma_k \cdot r_{k\Lambda})(\sigma_\Lambda \cdot r_{k\Lambda})/r^2_{k\Lambda} - (\sigma_k \cdot \sigma_\Lambda)$$

The dispersive part of the $\Lambda NN$ potential is given by,

$$V^{DS}_{\Lambda NN} = W_0 T_\pi^2(r_{i\Lambda}) T_\pi^2(r_{j\Lambda})[1 + \tfrac{1}{6}\sigma_\Lambda \cdot (\sigma_i + \sigma_j)] \quad (3)$$

$Y_\pi(r_{k\Lambda})$ and $T_\pi(r_{k\Lambda})$ are the usual Yukawa and tensor functions with pion mass, $\mu = 0.7$ fm$^{-1}$
Here $C_p$ and $W_0$ are $\Lambda NN$ interaction parameters.

The $\Lambda N$ and $\Lambda NN$ potential parameters for our preferred models are listed in Table I. $C_p$ and $W_0$ are the strength parameters of the two-pion and dispersive parts of the $\Lambda NN$ potential.

TABLE I : $\Lambda N$ and $\Lambda NN$ interaction parameters. Except for $\varepsilon$, all other quantities are in MeV.

| $\Lambda N$ | $\bar{V}$ | $V_\sigma$ | $\varepsilon$ | $C_p$ | $W_0$ |
|---|---|---|---|---|---|
| $\Lambda N1$ | 6.150 | 0.176 | 0.2 | 1.50 | 0.028 |
| $\Lambda N2$ | 6.110 | 0.000 | 0.0 | 1.50 | 0.028 |
| $\Lambda N3$ | 6.025 | 0.000 | 0.0 | 0.00 | 0.000 |

For $\Lambda\Lambda$ potential we use low-energy phase equivalent Nijmegen interactions represented by a sum of the three Gaussians[4,6,7],

$$V_{\Lambda\Lambda} = v^{(1)} \exp(-r^2/\beta^2_{(1)}) + \Upsilon v^{(2)} \exp(-r^2/\beta^2_{(2)}) + v^{(3)} \exp(-r^2/\beta^2_{(3)}) \quad (4)$$

Where the strength parameter $v^i$ and the range parameter $\beta_i$ are taken same as ref.[4]

We take the variational wave function to be of the form,

$$|\Psi_v\rangle = \left[1 + \sum_{i<j<k}(U_{ijk} + U^{TNI}_{ijk}) + \sum_{i<j,\Lambda} U_{ij,\Lambda} + \sum_{i<j} U^{LS}_{ij}\right] \prod_{i<j<k} f_c^{ijk} |\Psi_p\rangle \quad (5)$$

where, $|\Psi_p\rangle$ is the pair wave function[2,3] given by

$$|\Psi_p\rangle = S\prod_{i<j}(1+U_{ij}) S\prod_{i<\Lambda}(1+U_{i\Lambda}) |\Psi_J\rangle \quad (6)$$

The Jastrow wave function for lambda hypernuclei is given by,

$$|\Psi_J\rangle = [f_c^{\Lambda\Lambda} \prod_{i<j<k} f_c^{ijk} \prod_{i<\Lambda} f_c^{i\Lambda} \prod_{i<j} f_c^{ij}] |\Psi_{JT}\rangle |\varphi\rangle \quad (7)$$

where f's are the central correlation functions and |φ⟩ is an antisymmetric wave function of the lambda particle. $|\Psi_{JT}\rangle$ is the spin and isospin wavefuntion of the s-shell nucleus.

Variational Monte Carlo method is used to find the ground state energy and binding energy of different hypernuclear systems.

For double hypernuclear system, binding energy $B_{\Lambda\Lambda}$ is given by,

$$-B_{\Lambda\Lambda}(_{\Lambda\Lambda}^{A}Z) = E(_{\Lambda\Lambda}^{A}Z) - E(^{A-2}Z) \qquad (8)$$

The binding energy $B_\Lambda$ of a single hypernuclear system is given by,

$$-B_\Lambda(_\Lambda^A Z) = E(_\Lambda^A Z) - E(^{A-1}Z) \qquad (9)$$

## III. CALCULATION AND RESULTS :

We have already performed calculations on $_\Lambda^3 H$ and $_\Lambda^5 He$ with the three potential models ΛN1, ΛN2 and ΛN3[2,3]. The potential model , ΛN2 gave a binding energy for $_\Lambda^3 H$ as 0.13(00)[3] which agrees very well with experimental result. However ΛN3 could not give good result for $_\Lambda^3 H$. For $_\Lambda^5 He$, all the potential models have been tested earlier[2] and were found to be in good agreement with experimental result. Our results on $_\Lambda^4 H$ and earlier results for $_\Lambda^3 H$ and $_\Lambda^5 He$ for the three potential models are tabulated in Table II.

TABLE II : Results for $_\Lambda^4 H$ and $_\Lambda^5 He$ with ΛN2 and ΛN3. All the quantities are in Mev.

| Potential | Terms | $_\Lambda^3 H$ (Ref. 3) | $_\Lambda^5 He$ (Ref. 2) | $_\Lambda^4 H$ |
|---|---|---|---|---|
| ΛN1 | E | -2.56(01) | -30.89(03) | -10.55(01) |
|  | $B_\Lambda$ | **0.34(01)** | **3.17(03)** | **2.23(02)** |
| ΛN2 | E | -2.35(00) | -30.91(03) | -10.00(01) |
|  | $B_\Lambda$ | **0.13(00)** | **3.19(03)** | **1.68(02)** |
| ΛN3 | E |  | -30.85(01) | -9.02(01) |
|  | $B_\Lambda$ |  | **3.13(02)** | **0.70(02)** |
| *Expt.( $B_\Lambda$ )* |  | *0.13±0.05* | *3.12(02)* | *2.22(04)* |

From Table II , ΛN1 is found to be more appropriate for $_\Lambda^4 H$.

Then we calculated the binding energy for $_{\Lambda\Lambda}^5 H$ with our three preferred potential models ΛN1, ΛN2 and ΛN3.

TABLE III : Results for $_{\Lambda\Lambda}^{5}H$ with ΛN1, ΛN2 and ΛN3. All the quantities are in Mev.

| Potential | Terms | $_{\Lambda\Lambda}^{5}H$ |
|---|---|---|
| ΛN1 | E | -12.25(01) |
|  | **B**$_{\Lambda\Lambda}$ | **3.93(02)** |
| ΛN2 | E | -12.22(02) |
|  | **B**$_{\Lambda\Lambda}$ | **3.90(02)** |
| ΛN3 | E | -10.93(01) |
|  | **B**$_{\Lambda\Lambda}$ | **2.61(02)** |

## IV. CONCLUSION :

For the potentials ΛN1 and ΛN2 the binding energy for $_{\Lambda\Lambda}^{5}H$ is found to be consistent. For the double hypernuclear system $_{\Lambda\Lambda}^{5}H$, the third potential ΛN3 gives relatively smaller value of binding energy in comparison with ΛN1 and ΛN2. It indicates that the ΛNN potential is crucial in binding the hypernuclear system, $_{\Lambda\Lambda}^{5}H$.

For $_{\Lambda}^{4}H$, the first potential model in which the space exchange parameter ε is present, is found to be more close to experimental result. When both space exchange parameter and ΛNN potential are absent (third potential model ΛN3), the binding energy value for $_{\Lambda}^{4}H$ is found to differ considerably from experimental result. For $_{\Lambda\Lambda}^{5}H$ also the absence of space exchange parameter and three-body ΛNN potential seems to be crucial as it decreses the binding energy value by a substantial amount.

Therefore, we conclude that both space exchange part of ΛN potential and ΛNN potential plays important role in binding light lambda hypernuclei.

Bhupali Sharma acknowledges the financial support from UGC and the facilities at A. V. College, Guwahati, for the present work.